\documentclass[12pt,JStatPhys]{JArticle}

\usepackage{lettercount}
\usepackage{amsfonts}
\usepackage{epsf}

\begin{document}

\def\r{{\mathbf{r}}}
\def\R{{\mathbf{R}}}
\def\A{{\mathbf{A}}}
\def\M{{\mathbf{M}}}
\def\K{{\mathbf{K}}}
\def\G{{\mathbf{G}}}
\newcommand{\Pf}{\mathop{\mathrm{Pf}}}
\newcommand{\Tr}{\mathop{\mathrm{Tr}}}
\newcommand{\erf}{\mathop{\mathrm{erf}}}

\title{%
Two-dimensional Coulomb Systems in a disk with Ideal
Dielectric Boundaries }
\author{%
Gabriel T\'ellez%
{\def\baselinestretch{1}\footnote{%
Grupo de F\'{\i}sica Te\'orica de la Materia Condensada,
Departamento de F\'{\i}sica, Universidad de Los Andes, A.A.~4976,
Bogot\'a, Colombia; e-mail: gtellez@uniandes.edu.co
}
}
}
\date{}
\maketitle

\begin{abstract}
We consider two-dimensional Coulomb systems confined in a disk with
ideal dielectric boundaries. In particular we study the two-component
plasma in detail. When the coulombic coupling constant $\Gamma=2$ the
model is exactly solvable. We compute the grand-potential, densities
and correlations. We show that the grand-potential has a universal
logarithmic finite-size correction as predicted in previous works.
This logarithmic finite-size correction is also checked in another
solvable model: the one-component plasma.
\end{abstract}


\section{Introduction}

Solvable models of Coulomb systems have attracted attention for quite
some time. Very recently the two-dimensional two-component plasma, a
model of two species of point-particles with opposite charges $\pm q$
at inverse temperature $\beta=1/k_BT$, has been solved in its whole
range of stability $\Gamma:=\beta q^2<2$ by using a mapping of this
system into a sine-Gordon field theory.\citenote{SamajTravenec} With
this mapping the grand-partition function and other bulk properties of
the system can be computed exactly. Also some surface properties near
a metallic wall\citenote{JancoSamaj-metal} and an ideal dielectric
wall\citenote{Samaj-ideal-diel} have been investigated. However this
mapping into a sine-Gordon model does not give (yet) any information
on correlation functions.

When the coupling constant $\Gamma=2$ the corresponding sine-Gordon
model is at its free fermion point and additional information on the
system can be obtained. This fact is well known and much work has been
done on the two-component plasma at $\Gamma=2$ since the pioneer work
of Gaudin on this model.\citenote{Gaudin} In particular the
two-component plasma at $\Gamma=2$ has been studied in different
geometries: near a plane hard wall and a dielectric
interface\citenote{CornuJanco}, a metallic
wall\citenote{Forrester-metal-tcp}, in a disk with hard
walls\citenote{JancoManif} and in a disk with metallic
walls\citenote{JancoTellez-coulcrit}\dots.

However it was not until very recently that the special case of ideal
dielectric boundaries (that is Neumann boundary conditions imposed on
the electric potential) has been studied by Jancovici and
\v{S}amaj\citenote{JancoSamaj-diel} for a system near an infinite
plane wall or confined in a strip. The technical difficulty with this
kind of boundary conditions is that the two-component plasma must be
mapped into a four-component free Fermi field instead of a
two-component free Fermi field as in all other cases of boundary
conditions.

A very interesting result of reference~[\cite{JancoSamaj-diel}] is
that when the system is confined in a strip of width $W$ made of ideal
dielectric walls, the grand-potential per unit length exhibits a
universal finite-size correction equal to $\pi/24W$ which is the same
finite-size correction for a system confined in a strip made of ideal
conductor walls.\citenote{JancoTellez-coulcrit} These finite-size
corrections have been explained\citenote{JancoTellez-coulcrit} by
noting that if one disregards microscopic detail the grand-partition
function of a Coulomb system is the inverse of the partition function
of the Gaussian model.

Due to this analogy with the Gaussian field theory, a Coulomb
system in a confined geometry is expected to exhibit logarithmic
finite size-corrections, for instance in a disk of radius $R$ the
grand-potential should have a correction $(1/6)\ln R$. For the
analogy with the Gaussian field theory to be complete one should
impose conformally invariant boundary conditions to the electric
potential, for instance Dirichlet boundary conditions (ideal metallic
walls) or Neumann boundary conditions (ideal dielectric walls). The
case of Dirichlet boundary conditions was treated in
reference~[\cite{JancoTellez-coulcrit}].

In this paper we study the two-component plasma at $\Gamma=2$ in a
disk with Neumann boundary conditions. One of the main motivations for
this work is to show that the systems exhibits in fact the expected
logarithmic finite-size correction.

The outline of this paper is the following. In section 2, we present
the model and adapting the method of
reference~[\cite{JancoSamaj-diel}] we map the two-component plasma
into a four-component free Fermi field. In section 3, we compute the
grand-potential and its large-$R$ expansion. We also compute the
large-$R$ expansion of the free energy of the one-component plasma
which was solved some time ago.\citenote{Smith} In section 4, we
compute densities and correlation functions.

\section{The model}

The system is composed of two species of point-particles with
opposite charges $\pm q$. The particles are confined in a disk $D$ of
radius $R$. It will be very useful to work with the complex
coordinate $z=re^{i\phi}$ of a point $\r$.
The material outside the disk is supposed to be an ideal
dielectric. This imposes Neumann boundary conditions on the electric
potential: the interaction potential $v(\r,\r')$ between two charges
located at $\r$ and $\r'$ is the solution of Poisson equation
\begin{equation}\label{eq:Poisson}
\Delta_\r v(\r,\r')=-2\pi\delta(\r-\r')
\end{equation}
with Neumann boundary conditions.  However, it is
well-known\citenote{Jackson} that any solution of Poisson
equation~(\ref{eq:Poisson}) in a closed domain $D$ cannot satisfy
homogeneous Neumann boundary conditions $\partial_n v(\r,\r')=0$ for
$\r\in\partial D$,
since Gauss theorem implies that $\oint_{\partial D}
\partial_n v(\r,\r')\, dl = -2\pi$. A natural choice is to impose
inhomogeneous Neumann boundary conditions to the potential $\partial_n
v(\r,\r')=-2\pi/L$, with $L$ the perimeter of the domain $D$. In the
case of the disk of radius $R$ this gives 
\begin{equation}\label{eq:NeumannBC}
\partial_n v(\r,\r')=-1/R \quad \mathrm{for}\quad \r\in\partial D 
\end{equation}
This impossibility for the electric potential between pairs to obey
homogeneous Neumann boundary conditions 
is not a problem for a system
globally neutral in which the total electric potential will satisfy
homogeneous Neumann boundary conditions if the pair potential
satisfies~(\ref{eq:NeumannBC}). It should be noted that for an infinite
wall boundary it is possible for the electric potential between
pairs to obey homogeneous Neumann boundary
conditions.\citenote{JancoSamaj-diel}

The solution of Poisson equation~(\ref{eq:Poisson}) with boundary
conditions~(\ref{eq:NeumannBC}) in a disk can be obtained obtained by
usual methods of electrostatics (images, etc \dots). The solution is
\begin{equation}\label{eq:potential}
v(\r,\r')=-\ln\frac{|z-z'||R^2-z\bar{z}'|}{a^2 R}
\end{equation}
where $a$ is some irrelevant length scale.
It can be easily checked that solution~(\ref{eq:potential}) can also
be obtained as the limit of the boundary value problem where outside
the disk there is a dielectric with dielectric constant $\epsilon\to
0$ (up to some constant terms).\citenote{Smith}

It will be useful to introduce the image $\r^*$ of a point $\r$ by
$z^*=R^2/\bar{z}$. The electric potential between pairs can then be
written as
\begin{equation}
v(\r,\r')=-\ln\frac{|z-z'||z'^*-z||\bar{z}'|}{a^2 R}
\end{equation}
and can be interpreted as the potential in $\r$ due to a charge in
$\r'$ and a image charge with equal sign and magnitude located at
$\r'^*$. 

If the temperature is not high enough the two-dimensional
two-component plasma is not well defined, this is due to the collapse
between pairs of opposite sign. If the coupling constant
$\Gamma:=\beta q^2<2$ the system is stable. Here we will study the
case $\Gamma=2$. In order to avoid the collapse we will work initially
with two interwoven lattices $U$ and $V$. Positive particles with
complex coordinates $\{u_i\}$ live in the sites of lattice $U$ and
negative particles with complex coordinates $\{v_i\}$ live in the
sites of $V$. We shall work in the grand-canonical ensemble and will
only consider neutral configurations. The Boltzmann factor of a system
with $N$ positive particles and $N$ negatives particles at $\Gamma=2$
is
\begin{eqnarray}
e^{-\beta H_N}&= &a^{2N} \prod_{i=1}^{N} (R^2-|u_i|) (R^2-|v_i|)
\\
&&\times
\frac{\prod_{i<j}\left(|u_i-u_j||v_i-v_j||R^2-u_i \bar{u}_j| 
|R^2-v_i \bar{v}_j|\right)^2}%
{\prod_{i,j}\left(|u_i-v_j||R^2-u_i \bar{v}_j|\right)^2}
\nonumber
\end{eqnarray}
The first product corresponds to the self-energies of the particles
and the other terms to the interactions between pairs.
Introducing the images this can be rewritten as
\begin{eqnarray}
\lefteqn{e^{-\beta H_N}= a^{2N} 
\prod_{i}\left(\frac{R^2}{\bar{u}_i \bar{v}_i}\right)
\prod_{i=1}^{N} (u_i-u_i^*) (v_i-v_i^*)}
&&
\\
&\times&
\frac{\prod_{i<j}
(u_i-u_j)(u_i^*-u_j^*)(u_i-u_j^*)(u_i^*-u_j)
(v_i-v_j)(v_i^*-v_j^*)(v_i-v_j^*)(v_i^*-v_j)
}%
{\prod_{i,j}
(u_i-v_j)(u_i^*-v_j^*)(u_i-v_j^*)(u_i^*-v_j)}
\nonumber
\end{eqnarray}
By using Cauchy's double alternant formula 
\begin{equation}
\det\left(
\frac{1}{z_i-z'_j}
\right)_{(i,j)\in\{1,\dots,2N\}^2}
=(-1)^{N(2N-1)}
\frac{\prod_{i<j}(z_i-z_j)(z'_i-z'_j)}{\prod_{i,j}(z_i-z'_j)}
\end{equation}
with
\begin{equation}
z_{2i-1}=u_i\,,\ \ z_{2i}=u_i^*\,,\ \ 
z'_{2i-1}=v_i\,,\ \ z'_{2i}=v_i^*\ ,
\end{equation}
we find that the Boltzmann factor can be written as a $2N\times 2N$
determinant 
\begin{equation}
e^{-\beta H_N}=
(-1)^N \prod_i \left(\frac{a^2 R^2}{\bar{u}_i \bar{v}_i}
\right)
\,\det\left(
\begin{array}{cc}
\displaystyle
\frac{1}{u_i-v_j} & 
\displaystyle\frac{1}{u_i-v_j^*} \\
\displaystyle \frac{1}{u_i^*-v_j} & 
\displaystyle \frac{1}{u_i^*-v_j^*} \\
\end{array}
\right)
\end{equation}
Introducing the factors $Ra/\bar{u}_i$ into the last $N$ rows of the
determinant and the factors $Ra/\bar{v}_i$ into the last $N$ columns,
we finally arrive at the expression
\begin{equation}\label{eq:Boltzmann}
e^{-\beta H_N}=
(-1)^N 
\,\det\left(
\begin{array}{cc}
\displaystyle
\frac{a}{u_i-v_j} & 
\displaystyle\frac{aR}{u_i \bar{v}_j-R^2} \\
\displaystyle \frac{aR}{R^2-\bar{u}_i v_j} & 
\displaystyle \frac{a}{\bar{v}_j-\bar{u}_i} \\
\end{array}
\right)
\end{equation}
The grand-canonical partition function with position dependent
fugacities $\zeta(u)$ and $\zeta(v)$ for positive and negative
particles reads
\begin{equation}\label{eq:grand-partition}
\Xi=1+\sum_{N=1}^\infty
\sum_{{u_1<\cdots<u_N \in U}\atop
{v_1<\cdots<v_N \in V}}
\left(\prod_{i=1}^N \zeta(u_i) \zeta(v_i)\right)
e^{-\beta H_N}
\end{equation}

We shall now closely follow reference~[\cite{JancoSamaj-diel}] to show
that the grand-partition function can be written as a ratio of two
Pfaffians.  Using the same notations as in
reference~[\cite{JancoSamaj-diel}], let us introduce a couple of
Grassmann anticommuting variables $(\psi^1_u,\psi^2_u)$ for each site
$u\in U$ and similar Grassmann variables for each site in $V$. The
Grassmann integral for any anti-symmetric matrix $\A$
\begin{equation}
Z_0=\int d\theta\ \exp\left(\frac{1}{2} {}^t\theta \A \theta
\right)
\end{equation}
with
${}^t\theta=(\dots,\psi^1_u,\psi^2_u,\dots,\psi^1_v,\psi^2_v,\dots)$
and $d\theta=\prod_v d\psi_v^2\psi_v^1 \prod_u \psi_u^2\psi_u^1$ is
the Pfaffian of the matrix $\A$
\begin{equation}
Z_0=\Pf(\A)
\end{equation}
Let us denote the average of a quantity ${\cal O}$ by
\begin{equation}
\left<{\cal O}\right>=\frac{1}{Z_0}
\int d\theta\ {\cal O} \ 
\exp\left(\frac{1}{2} {}^t\theta \A \theta \right)
\end{equation}
It is well known that\citenote{ItzyksonDrouffe}
\begin{equation}
\left<\theta_i\theta_j\right>=(A^{-1})_{ji}
\end{equation}
For our purposes let us define 
the matrix $\A$ as having inverse elements 
\begin{mathletters}
\begin{eqnarray}
(A^{-1})^{\alpha\beta}_{uu'}&=&0\\
(A^{-1})^{\alpha\beta}_{vv'}&=&0\\
(A^{-1})^{\alpha\beta}_{uv}&=&
\left(
\begin{array}{cc}
\displaystyle
\frac{a}{u-v} & 
\displaystyle\frac{aR}{u \bar{v}-R^2} \\
\displaystyle \frac{aR}{R^2-\bar{u} v} & 
\displaystyle \frac{1}{\bar{v}-\bar{u}} \\
\end{array}
\right)\\
(A^{-1})^{\alpha\beta}_{vu}&=&
\left(
\begin{array}{cc}
\displaystyle
\frac{a}{v-u} & 
\displaystyle\frac{aR}{v \bar{u}-R^2} \\
\displaystyle \frac{aR}{R^2-\bar{v} u} & 
\displaystyle \frac{a}{\bar{u}-\bar{v}} \\
\end{array}
\right)
\end{eqnarray}
\end{mathletters}
The indexes $(\alpha,\beta)\in\{1,2\}^2$ label the rows and columns
respectively. The matrix $\A$ is clearly anti-symmetric as required. We
now introduce the anti-symmetric matrix $\M$
\begin{mathletters}
\begin{eqnarray}
M^{\alpha\beta}_{uu'}&=&
\delta_{uu'} 
\left(
\begin{array}{cc}
\displaystyle
0& \zeta(u)\\
-\zeta(u) & 0 \\
\end{array}
\right)
\\
M^{\alpha\beta}_{vv'}&=&
\delta_{vv'} 
\left(
\begin{array}{cc}
\displaystyle
0& \zeta(v)\\
-\zeta(v) & 0 \\
\end{array}
\right)
\\
M^{\alpha\beta}_{uv}&=&0\\
M^{\alpha\beta}_{vu}&=&0
\end{eqnarray}
\end{mathletters}
The Grassmann integral 
\begin{equation}
Z=\int d\theta\ \exp\left[\frac{1}{2} {}^t\theta(\A+\M)\theta\right]
\end{equation}
is equal to 
\begin{equation}
Z=\Pf(\A+\M)
\end{equation}
The ratio $Z/Z_0$ can be expanded in powers of the fugacities as
\begin{equation}\label{eq:Z/Zo}
\frac{Z}{Z_0}
=1+\sum_{N=1}^\infty
\sum_{{u_1<\cdots<u_N \in U}\atop
{v_1<\cdots<v_N \in V}}
\left(\prod_{i=1}^N \zeta(u_i) \zeta(v_i)\right)
\left<\prod_{i=1}^N \left(
\psi^1_{u_i}\psi^2_{u_i}\psi^1_{v_i}\psi^2_{v_i}
\right)
\right>
\end{equation}
Using Wick theorem for fermions we find that
\begin{equation}\label{eq:average-psi}
\left<\prod_{i=1}^N \left(
\psi^1_{u_i}\psi^2_{u_i}\psi^1_{v_i}\psi^2_{v_i}
\right)
\right>
= (-1)^N
\,\det\left(
\begin{array}{cc}
\displaystyle
\frac{a}{u_i-v_j} & 
\displaystyle\frac{aR}{u_i \bar{v}_j-R^2} \\
\displaystyle \frac{aR}{R^2-\bar{u}_i v_j} & 
\displaystyle \frac{a}{\bar{v}_j-\bar{u}_i} \\
\end{array}
\right)
\end{equation}
Comparing equations~(\ref{eq:Z/Zo}) and~(\ref{eq:average-psi}) with
equations~(\ref{eq:grand-partition}) and~(\ref{eq:Boltzmann}) we
immediately conclude that the grand-canonical partition of the Coulomb
system $\Xi$ is the ratio of two Pfaffians
\begin{equation}
\Xi=\frac{Z}{Z_0}=\frac{\Pf(\A+\M)}{\Pf(\A)}
\end{equation}
Using the fact that the Pfaffian is the square root of the
determinant we can write the grand-potential as
\begin{equation}\label{eq:grand-potential}
\beta \Omega=-\ln\Xi=
-\frac{1}{2} \ln\det(1+\K)=
-\frac{1}{2} \Tr\ln(1+\K)
\end{equation}
where the matrix $\K$ is $\K=\M\A^{-1}$ and has matrix elements
\begin{equation}
K_{ss'}^{\alpha\beta}(\r,\r')=
\delta_{s,-s'}\zeta_s(\r)
\left(
\begin{array}{cc}
\displaystyle
\frac{aR}{R^2-\bar{z}z'} & 
\displaystyle\frac{a}{\bar{z}'-\bar{z}}\\
\displaystyle \frac{a}{z'-z} &
\displaystyle \frac{aR}{R^2-z\bar{z}'}\\
\end{array}
\right)
\end{equation}
We have introduced the notation $u=(\r,+)$, $v=(\r,-)$,
$\zeta(u)=\zeta_+(\r)$, $\zeta(v)=\zeta_-(\r)$ and
$(s,s')\in\{-,+\}^2$. 

\section{The grand-potential}

\subsection{Formal expression of the grand-potential}

To compute explicitly the grand-potential from
equation~(\ref{eq:grand-potential}) we must find the eigenvalues of
$\K$. From now on we will consider the continuum limit where the
spacing of the lattices $U$ and $V$ goes to zero. In this limit it is
natural to work with the re-scaled fugacity\citenote{CornuJanco}
$m=2\pi a \zeta/S$ where $S$ is the area of a lattice cell. Also in
this limit some bulk quantities will be divergent, because of the
collapse of particle of opposite sign, and must be
cutoff. Correlations in contrast have a well defined continuum limit.

Let $\{\psi_s^{(\alpha)}(\r)\}_{s=\pm;\, \alpha=1,2}$ be the
eigenvectors of $m^{-1}\K$ and $1/\lambda$ the corresponding
eigenvalues. The eigenvalue problem for $\K$ is the set of integral
equations
\begin{mathletters}\label{eq:integralK}
\begin{eqnarray}
\label{eq:integralK1}
\frac{\lambda}{2\pi}
\int_D d\r' \left[
\frac{R}{R^2-\bar{z}z'}\psi_{-s}^{(1)}(\r')+
\frac{1}{\bar{z}'-\bar{z}}\psi_{-s}^{(2)}(\r')
\right]&=&\psi_{s}^{(1)}(\r)\\
\label{eq:integralK2}
\frac{\lambda}{2\pi}
\int_D d\r' \left[
\frac{1}{z'-z}\psi_{-s}^{(1)}(\r')+
\frac{R}{R^2-z\bar{z}'}\psi_{-s}^{(2)}(\r')
\right]&=&\psi_{s}^{(2)}(\r)
\end{eqnarray}
\end{mathletters}
These integral equations~(\ref{eq:integralK}) can be transformed into
differential equations plus some boundary conditions using the
well-known identities
\begin{equation}
\frac{\partial}{\partial z} \frac{1}{\bar{z}-\bar{z}'}=
\frac{\partial}{\partial \bar{z}} \frac{1}{z-z'}=
\pi\delta(\r-\r')\,.
\end{equation}
Applying $\partial_z$ to equation~(\ref{eq:integralK1}) and
$\partial_{\bar{z}}$ to equation~(\ref{eq:integralK2}) yields
\begin{mathletters}\label{eq:differentialK}
\begin{eqnarray}
\label{eq:differentialK1}
-\frac{\lambda}{2} \psi_{-s}^{(2)}(\r)&=&\partial_z\psi_{s}^{(1)}(\r)
\\
\label{eq:differentialK2}
-\frac{\lambda}{2} \psi_{-s}^{(1)}(\r)&=&
\partial_{\bar{z}}\psi_{s}^{(2)}(\r)
\end{eqnarray}
\end{mathletters}
These differential equations~(\ref{eq:differentialK}) can be combined
into the Laplacian eigenvalue problem
\begin{equation}\label{eq:Laplace}
\Delta \psi_{s}^{(\alpha)}=\lambda^2  \psi_{s}^{(\alpha)}
\end{equation}
which must be complemented with the following boundary conditions. If
$\r=\R$ is on the boundary, $z=Re^{i\phi}$, it can be easily seen from
integral equations~(\ref{eq:integralK}) that
\begin{equation}\label{eq:BCforK}
\psi_{s}^{(1)}(\R)+e^{i\phi}\psi_{s}^{(2)}(\R)=0
\end{equation}
An elementary solution of equation~(\ref{eq:Laplace}) in the present
disk geometry is
\begin{equation}
\psi_{s}^{(2)}(\r)=A_s e^{i\ell\phi} I_\ell(\lambda r)
\end{equation}
with, from equation~(\ref{eq:differentialK2}),
\begin{equation}
\psi_{-s}^{(1)}(\r)=-A_s e^{i(\ell+1)\phi} I_{\ell+1}(\lambda r)
\end{equation}
where $I_{\ell}$ is a modified Bessel functions of order
$\ell$. Boundary conditions~(\ref{eq:BCforK}) yield the following
homogeneous linear system for the coefficients $A_s$
\begin{mathletters}
\begin{eqnarray}
-A_- I_{\ell+1}(\lambda R) + A_+ I_\ell (\lambda R)&=&0\\
A_- I_{\ell}(\lambda R) - A_+ I_{\ell+1} (\lambda R)&=&0
\end{eqnarray}
\end{mathletters}
For this system to have non trivial solutions its determinant must
vanish. This gives the equation for the eigenvalue $\lambda$
\begin{equation}\label{eq:for-lambda}
I_{\ell+1}(\lambda R)^2-
I_{\ell}(\lambda R)^2=0
\end{equation}
From equation~(\ref{eq:grand-potential}) the grand-potential then reads
\begin{equation}\label{eq:Omega-prod}
\beta\Omega=-\frac{1}{2}\sum_{\ell=-\infty}^{+\infty}
\ln\prod_\lambda \left(1+\frac{m}{\lambda}\right)
=-
\sum_{\ell=0}^{+\infty}
\ln\prod_\lambda \left(1+\frac{m}{\lambda}\right)
\end{equation}
where the product runs for all $\lambda$ solution of
equation~(\ref{eq:for-lambda}). 
The last equality in equation~(\ref{eq:Omega-prod}) comes from
noticing that a change $\ell\to-\ell-1$ leave
equation~(\ref{eq:for-lambda}) invariant.
To evaluate the product in
equation~(\ref{eq:Omega-prod}), let us introduce the analytic function
for $\ell$ positive
\begin{equation}
f_\ell(z)=\left(
I_{\ell}(zR)^2-I_{\ell+1}(zR)^2
\right)
\left(
\left(
\frac{2}{zR}
\right)^\ell
\ell!
\right)^2
\end{equation}
The zeros of this function are the eigenvalues $\lambda$ and it can
be checked that $f'_\ell(z)/(zf_\ell(z))\to 0$ as $z\to\infty$, so
this function can be factorized as a Weierstrass product
\begin{equation}\label{eq:Weierstrass}
f_\ell(z)=f_\ell(0)e^{f'_\ell(0)z/f_\ell(0)}
\prod_{\lambda} \left(1-\frac{z}{\lambda}\right) e^{z/\lambda}
\end{equation}
This function satisfies $f_\ell(0)=1$, $f'_\ell(0)=0$, and
$f_\ell(z)=f_\ell(-z)$ so its zeros come in pairs of opposite sign
and as a consequence the exponential factors in Weierstrass
product~(\ref{eq:Weierstrass}) cancels out. We finally have
\begin{equation}
f_\ell(z)=\prod_\lambda (1-z/\lambda)
\end{equation}
where the product runs over all $\lambda$ solution of
equation~(\ref{eq:for-lambda}). 

Then the grand-potential can finally be written as
\begin{eqnarray}\label{eq:Omega-final}
\beta\Omega&=&-\sum_{l=0}^{+\infty}\ln f_\ell(-m)\\
&=&-\sum_{l=0}^{+\infty}
\ln\left[
\left(\frac{2}{mR}\right)^{2\ell}
(\ell!)^2
\left(
I_\ell(mR)^2-I_{\ell+1}(mR)^2
\right)
\right]
\nonumber
\end{eqnarray}
The above expression is divergent and must be cutoff to a
$\ell_\mathrm{max}=R/\sigma$ where $\sigma$ is the ratio of the
particles.\citenote{CornuJanco} 

It is interesting to notice that the grand-potential can be written as
the sum of two terms
\begin{equation}\label{eq:Omega-decompo}
\Omega=\frac{1}{2}
\left[\Omega_{\mathrm{ideal\atop cond}}(m)+\Omega_{\mathrm{ideal\atop
cond}}(-m)
\right]
\end{equation}
where
\begin{equation}
\beta\Omega_{\mathrm{ideal\atop cond}}(m)=
-2
\sum_{l=0}^\infty 
\ln\left[
\left(\frac{2}{mR}\right)^{\ell}
\ell!
\left(
I_\ell(mR)+I_{\ell+1}(mR)
\right)
\right]
\end{equation}
is the grand-potential of a two-component plasma at $\Gamma=2$
confined in a disk with ideal {\em conductor\/}
boundaries\citenote{JancoTellez-coulcrit} and
\begin{equation}
\beta\Omega_{\mathrm{ideal\atop cond}}(-m)=
-2\sum_{l=0}^\infty 
\ln\left[
\left(\frac{2}{mR}\right)^{\ell}
\ell!
\left(
I_\ell(mR)-I_{\ell+1}(mR)
\right)
\right]
\end{equation}
is formally the grand-potential of the two-component plasma with ideal
conductor boundaries but with the sign of the fugacity changed (which
of course does not correspond to any physical system).
This interesting decomposition of the grand-potential also exist in
the strip geometry.\citenote{JancoSamaj-diel}

\subsection{Finite-size corrections}

We now compute the large-$R$ expansion of the grand
potential~(\ref{eq:Omega-final}). 
First using decomposition~(\ref{eq:Omega-decompo}) we can use the
results of reference~[\cite{JancoTellez-coulcrit}] for the expansion
of the grand-potential with ideal conductor boundaries
\begin{equation}
\beta\Omega_{\mathrm{ideal\atop cond}}(m)
=
-\beta p_b\pi R^2 + \beta\gamma_c 2\pi R +
\frac{1}{6} \ln (mR)
+O(1)
\end{equation}
where the bulk pressure is
\begin{equation}
\beta p_b=\frac{m^2}{2\pi}\left(\ln\frac{2}{m\sigma}+1\right)
\end{equation}
and the surface contribution with ideal conductor boundaries is
\begin{equation}
\beta\gamma_c=m 
\left(
-\frac{1}{2\pi}\ln\frac{2}{m\sigma} - \frac{1}{2\pi}+
\frac{1}{4}
\right)
\end{equation}
The second contribution to the grand-potential can be written as
\begin{equation}
\frac{1}{2}
\beta\Omega_{\mathrm{ideal\atop cond}}(-m)
=
-\sum_{\ell=0}^{R/\sigma}
\ln \left[1-\frac{I_{\ell+1}(mR)}{I_\ell(mR)}\right]
+\frac{1}{2}\beta\Omega_{\mathrm{hard}}
\end{equation}
where
\begin{equation}
\beta\Omega_{\mathrm{hard}}=
-2
\sum_{\ell=0}^{R/\sigma} 
\ln\left[\ell!
\left(\frac{2}{mR}\right)^\ell
I_\ell(mR)\right]
\end{equation}
is the grand-potential of a two-component plasma in a disk with hard
wall boundaries.\citenote{JancoManif} The asymptotic expansion of this
term was computed in reference~[\cite{JancoManif}] and reads
\begin{equation}
\beta\Omega_{\mathrm{hard}}=
-\beta p_b \pi R^2
+\beta\gamma_h 2\pi R
+\frac{1}{6} \ln(mR)
+O(1)
\end{equation}
with the surface contribution for hard walls
\begin{equation}
\beta \gamma_h= m \left(\frac{1}{4}-\frac{1}{2\pi}\right)
\end{equation}
Finally, we only need to compute the asymptotic expansion of
\begin{equation}
S=-\sum_{\ell=0}^{R/\sigma}
\ln \left[1-\frac{I_{\ell+1}(mR)}{I_\ell(mR)}\right]
\end{equation}
This can be done with Debye asymptotic expansions for the Bessel
functions.
First let us write $S$ as
\begin{equation}
S=-\sum_{\ell=0}^{R/\sigma}
\ln\left[1-\frac{I_\ell'(mR)}{I_\ell(mR)}+\frac{\ell}{mR}\right]
\end{equation}
The Debye asymptotic expansion for $\ln I_\ell$
is\citenote{JancoManif,WatsonBessel} 
\begin{eqnarray}
\ln I_\ell(mR)&=&-\frac{1}{2}\ln(2\pi)-\frac{1}{4}\ln((mR)^2+\ell^2)
\nonumber\\
&&+\left((mR)^2+\ell^2\right)^{1/2}
-\ell\ln \frac{\ell+\sqrt{\ell^2+(mR)^2}}{mR}
\\
&&+O\left((mR)^2+\ell^2)^{-1/2}\right)
\nonumber
\end{eqnarray}
Therefore
\begin{eqnarray}\label{eq:DebyeExpansion}
\frac{I_\ell'(mR)}{I_\ell(mR)}&=&
-\frac{1}{2}\frac{mR}{(mR)^2+\ell^2}
+\frac{mR}{\sqrt{(mR)^2+\ell^2}}
+\frac{\ell}{mR}
\nonumber\\
&&
-\frac{\ell mR}{\left[(mR)^2+\ell^2\right]^{1/2}
\left[\ell+\sqrt{(mR)^2+\ell^2}\right]}
\\
&&+O\left((mR)^2+\ell^2)^{-1/2}\right)
\nonumber
\end{eqnarray}
Using expansion~(\ref{eq:DebyeExpansion}) and using Euler-MacLaurin
formula for the sum over~$\ell$
\begin{equation}
\sum_{\ell=0}^{\ell_\mathrm{max}}
f(\ell)= 
\int_{0}^{\ell_\mathrm{max}} f(x) \,dx
+
\frac{1}{2}
\left[f(0)+f(\ell_\mathrm{max})\right] + \cdots
\end{equation}
we finally find
\begin{equation}
S=mR +\frac{mR}{2} \ln \frac{2}{m\sigma} + O(1)
\end{equation}
We notice that the divergent (when the cutoff $\sigma\to 0$) surface
contribution from $\beta\Omega_\mathrm{ideal\atop cond}(m)$ is
canceled by the contribution from $S$ giving for the surface tension
the already known\citenote{Samaj-ideal-diel,JancoSamaj-diel} finite
expression
\begin{equation}
\beta \gamma_d= \frac{m}{4}
\end{equation}
Putting all terms together
\begin{equation}
\beta\Omega=
-\beta p_b \pi R^2
+\beta\gamma_d 2\pi R
+\frac{1}{6} \ln (mR) + O(1)
\end{equation}
The grand-potential has the expected $(1/6) \ln (mR)$ finite-size
correction.

\subsubsection*{The one-component plasma}

As a complement to the above study of the finite-size corrections, in
this section we consider another solvable model of Coulomb system, the
two-dimensional one-component plasma at
$\Gamma=2$.\citenote{JancoAlastuey,JancoOCP} This systems is composed
of $N$ particles with charge $q$ living in a neutralizing uniform
background. The one-component plasma in a disk with ideal dielectric
boundaries was solved by Smith.\citenote{Smith} The canonical
partition function reads
\begin{eqnarray}\label{eq:Zocp}
Z&=&
\left(\frac{\pi Ra}{\lambda_{\mathrm{th}}^2}
\right)^N
e^{3N^2/4}
N^{-N(N+1)/2}
\\
&&\times
\prod_{l=1}^N
\left(
\gamma(l,N)-N^{-(2N+1-2l)}
\gamma(2N+1-l,N)\right)
\nonumber
\end{eqnarray}
with $\gamma(s,x)=\int_0^x t^{s-1} e^{-t} \,dt$ the incomplete
gamma function and $\lambda_{\mathrm{th}}=h/\sqrt{2\pi k_B T}$ is the
thermal wavelength of the particles.

We want to study the large-$R$ expansion of the free energy of the
one-component plasma.
The free energy can be written as
\begin{equation}\label{eq:ocp-free-energy}
\beta F=\beta F_{\mathrm{hard}}-
\sum_{n=0}^{N-1}
\ln\left[1-
N^{-(2N-1-2n)}
\frac{\gamma(2N-n,n)}{\gamma(1+n,N)}
\right]
\end{equation}
where 
\begin{eqnarray}
\beta F_{\mathrm{hard}}&=&-3N^2/4-N
\ln\left(\frac{\pi Ra}{\lambda_{\mathrm{th}}^2}
\right)+
(N(N+1)/2)\ln N
\\
&&
-\sum_{n=1}^{N} \ln\gamma(n,N)
\nonumber
\end{eqnarray}
is the free energy of a one-component plasma in a disk with hard walls
boundaries.\citenote{JancoAlastuey} The finite-size expansion of this
terms is\citenote{JancoManif}
\begin{equation}\label{eq:ocp-free-energy-expansion-hard}
\beta F_{\mathrm{hard}}=\beta f \pi R^2 + \beta
\gamma_{\mathrm{ocp\atop hard}} 2\pi R + \frac{1}{6} \ln \left[
(\pi n)^{1/2} R\right]+O(1)
\end{equation}
with the bulk free energy per unit ``volume'' (surface)
\begin{equation}
\beta f = \frac{n}{2} \ln \left[\frac{n\lambda_{\mathrm{th}}^4}{2\pi^2
a^2}\right]
\end{equation}
and the ``surface'' (perimeter) contribution
\begin{equation}
\beta \gamma_{\mathrm{ocp\atop hard}}=
-\sqrt{\frac{n}{2\pi}}
\int_0^\infty \ln \frac{1+\erf(y)}{2}\,dy
\end{equation}
$n=N/(\pi R^2)$ is the density and $\erf(y)=(2/\sqrt{\pi})\int_0^y
e^{-x^2}\,dx$ is the error function.

The expansion of the remaining term in
equation~(\ref{eq:ocp-free-energy}) can be obtained with the following
uniform asymptotic expansions for the incomplete gamma function 
\begin{mathletters}
\begin{eqnarray}
\gamma(n+1,N)&=&\frac{n!}{2} \left[1+\erf\left(\frac{N-n}{\sqrt{2N}}
\right)+O\left(1/\sqrt{N}\right)\right]
\\
\gamma(2N-n,N)&=&\frac{(2N-n-1)!}{2}
\nonumber\\
&&\times 
\left[1+\erf\left(\frac{n-N}{\sqrt{2N}}\right)
+O\left(1/\sqrt{N}\right)\right]
\end{eqnarray}
\end{mathletters}
These expansions are valid when $N-n$ is of order $\sqrt{N}$ and the
corresponding terms in the sum~(\ref{eq:ocp-free-energy}) are the ones
that give a relevant contribution to $\beta F$.  Also for $n$ such
that $N-n$ is of order $\sqrt{N}$ using Stirling formula we have the
expansion for the factorials
\begin{mathletters}
\begin{eqnarray}
(2N-n-1)!&=&N!\,N^{N-n-1}\left[1+O\left(1/\sqrt{N}\right)\right]\\
n!&=&N!\,N^{n-N}\left[1+O\left(1/\sqrt{N}\right)\right]
\end{eqnarray}
\end{mathletters}
Finally replacing the sum in equation~(\ref{eq:ocp-free-energy}) by
an integral we find
\begin{equation}
\beta F= \beta F_{\mathrm{hard}}
- \sqrt{2N}
\int_0^\infty
\ln\frac{2\erf(y)}{1+\erf(y)}\ dy
+O(1)
\end{equation}
Putting this last result together with the
expansion~(\ref{eq:ocp-free-energy-expansion-hard}) for the hard wall
case we find
\begin{equation}\label{eq:ocp-free-energy-expansion}
\beta F_{\mathrm{hard}}=\beta f \pi R^2 + \beta
\gamma_{\mathrm{ocp\atop diel}} 2\pi R + \frac{1}{6} \ln \left[
(\pi n)^{1/2} R\right]+O(1)
\end{equation}
with
\begin{equation}
\beta\gamma_{\mathrm{ocp\atop diel}}=
-\sqrt{\frac{n}{2\pi}}\,
\int_0^\infty
\ln \erf(y) \ dy
\end{equation}
For this model we find again the expected logarithmic finite-size
correction.

\section{Density and correlations}

\subsection{Green functions}

We return to the study of the two-component plasma. We are now
interested in the density and correlations functions.
These can be obtained with the Green function 
\begin{equation}\label{eq:defG}
\G=\frac{1}{2\pi a\zeta}\frac{\K}{1+\K}
\end{equation}
as explained in reference~[\cite{JancoSamaj-diel}]. The density
$n_s(\r)$ of particles of sign $s$ is
\begin{equation}\label{eq:density-def}
n_s(\r)=\frac{m}{2}\,\sum_\alpha G_{ss}^{\alpha\alpha}(\r,\r)
\end{equation}
and the truncated two-body density is
\begin{equation}\label{eq:correlations-def}
n_{s_1 s_2}^{(2)T}(\r_1,\r_2)=
-\frac{m^2}{2}
\sum_{\alpha_1\alpha_2}
G_{s_1 s_2}^{\alpha_1\alpha_2}(\r_1,\r_2)
G_{s_2 s_1}^{\alpha_2\alpha_1}(\r_2,\r_1)
\end{equation}
From its definition~(\ref{eq:defG}) the Green functions obey the
integral equations
\begin{mathletters}\label{eq:Gintegral}
\begin{eqnarray}
G_{ss}(\r_1,\r_2)+\frac{m}{2\pi}
\int d\r
\left(
\begin{array}{cc}
\frac{R}{R^2-\bar{z}_1 z} & \frac{1}{\bar{z}-\bar{z}_1} \\
\frac{1}{z-z_1}& \frac{R}{R^2-z_1\bar{z}} 
\end{array}
\right)
G_{-ss}(\r,\r_2)&=&0\\
G_{-ss}(\r_1,\r_2)+\frac{m}{2\pi}
\int d\r
\left(
\begin{array}{cc}
\frac{R}{R^2-\bar{z}_1 z} & \frac{1}{\bar{z}-\bar{z}_1} \\
\frac{1}{z-z_1}& \frac{R}{R^2-z_1\bar{z}} 
\end{array}
\right)
G_{ss}(\r,\r_2)&=&\nonumber\\
=\frac{1}{2\pi}
\left(
\begin{array}{cc}
\frac{R}{R^2-\bar{z}_1 z_2} & \frac{1}{\bar{z}_2-\bar{z}_1} \\
\frac{1}{z_2-z_1}& \frac{R}{R^2-z_1\bar{z}_2} 
\end{array}
\right)
\end{eqnarray}
\end{mathletters}
These integral equations can be transformed into the differential
equations
\begin{mathletters}\label{eq:differentialG}
\begin{eqnarray}\label{eq:differentialG1}
G_{-ss}(\r_1,\r_2)-\frac{2}{m}
\left(
\begin{array}{cc}
0 & \partial_{\bar{z}_1}\\
\partial_{z_1} & 0 \\
\end{array}
\right)
G_{ss}(\r_1,\r_2)
&=&0
\\
\label{eq:differentialG2}
\left(
\begin{array}{cc}
0 & \partial_{\bar{z}_1}\\
\partial_{z_1} & 0 \\
\end{array}
\right)
G_{-ss}(\r_1,\r_2)
-\frac{m}{2} G_{ss}(\r_1,\r_2)&=&
-\frac{1}{2}
\delta(\r_1-\r_2)\mathbb{I}
\end{eqnarray}
\end{mathletters}
where $\mathbb{I}$ is the $2\times2$ unit matrix. These equations can
be combined into
\begin{equation}\label{eq:HelmoltzG}
\Delta_{\r_1} G_{ss}(\r_1,\r_2)-m^2
G_{ss}(\r_1,\r_2)=-m\delta(\r_1-\r_2)
\mathbb{I}
\end{equation}
The boundary conditions can be obtained from the integral
equations~(\ref{eq:Gintegral}). If $\r_1=\R$, $z_1=R e^{i\phi_1}$, is
on the boundary then from the integral equations~~(\ref{eq:Gintegral})
it can be seen that
\begin{mathletters}\label{eq:G-bc}
\begin{eqnarray}
G_{ss'}^{11}(\R,\r_2)+e^{i\phi_1} G_{ss'}^{21}(\R,\r_2)&=&0\\
G_{ss'}^{12}(\R,\r_2)+e^{i\phi_1} G_{ss'}^{22}(\R,\r_2)&=&0
\end{eqnarray}
\end{mathletters}
For the present disk geometry we look for a solution of
equation~(\ref{eq:HelmoltzG}) as a Fourier series in $\phi_1$. The
solution for $G_{ss}^{11}$ and $G_{ss}^{21}$ can be written as
\begin{mathletters}\label{eq:G-ctes}
\begin{eqnarray}
G_{ss}^{11}(\r_1,\r_2)
&=&
\frac{1}{2\pi}
\sum_{\ell\in\mathbb{Z}} 
e^{i\ell\phi_1}
\left[
me^{-i\ell\phi_2}
I_\ell(mr_<)K_\ell(mr_>) 
\right.\nonumber\\
&&
\left.\qquad\qquad\qquad
+A_\ell I_\ell(mr_1)I_\ell(mr_2)\right]
\\
G_{ss}^{21}(\r_1,\r_2)
&=&
\frac{1}{2\pi}\sum_{\ell\in\mathbb{Z}} e^{i\ell\phi_1} 
B_\ell I_\ell(mr_1)
\end{eqnarray}
\end{mathletters}
where $K_\ell$ is a modified Bessel function of order $\ell$,
$r_<=\min(r_1,r_2)$, $r_>=\max(r_1,r_2)$ and $A_\ell$ and $B_\ell$ are
constants (with respect to $\r_1$) of integration that will be
determined by the boundary conditions~(\ref{eq:G-bc}).
From equation~(\ref{eq:differentialG1}) we have
\begin{mathletters}
\begin{eqnarray}
G_{-ss}^{11}(\r_1,\r_2)&=&\frac{2}{m}\partial_{\bar{z}_1}
G_{ss}^{21}(\r_1,\r_2)\\ 
G_{-ss}^{21}(\r_1,\r_2)&=&\frac{2}{m}\partial_{z_1}G_{ss}^{11}(\r_1,\r_2)
\end{eqnarray}
\end{mathletters}
Therefore, using equations~(\ref{eq:G-ctes}) we have, for $r_1>r_2$,
\begin{mathletters}
\begin{eqnarray}
G_{-ss}^{21}(\r_1,\r_2)
&=&
\frac{1}{2\pi}
\sum_{\ell\in\mathbb{Z}} 
e^{i(\ell-1)\phi_1} I_\ell(mr_2)
\left[-me^{-i\ell\phi_2} K_{\ell-1}(mr_1) 
\right.\nonumber\\
&&
\left.\qquad\qquad\qquad
+
A_\ell I_{\ell-1}(mr_1)\right]
\\
G_{ss}^{11}(\r_1,\r_2)
&=&
\frac{1}{2\pi}\sum_{\ell\in\mathbb{Z}} e^{i\ell\phi_1} 
B_{\ell-1}I_{\ell-1}(mr_1)
\end{eqnarray}
\end{mathletters}
Using the boundary conditions~(\ref{eq:G-bc}) we find the following
linear system of equations for $A_\ell$ and $B_\ell$
\begin{mathletters}
\begin{eqnarray}
A_\ell I_\ell(mr_2) I_\ell + 
B_{\ell-1} I_{\ell-1}&=&-me^{-i\ell\phi_2}K_\ell I_\ell(mr_2)
\\
A_\ell I_\ell(mr_2) I_{\ell-1} + 
B_{\ell-1} I_\ell&=&me^{-i\ell\phi_2}K_{\ell-1}I_\ell(mr_2)
\end{eqnarray}
\end{mathletters}
where $I_\ell=I_\ell(mR)$ and similar definitions for the other Bessel
functions without argument. The solution of this linear system is
\begin{eqnarray}
A_\ell&=&-me^{-i\ell\phi_2}\
\frac{K_\ell I_\ell+K_{\ell-1}I_{\ell-1}}{I_\ell^2-I_{\ell-1}^2}
\\
B_{\ell-1}&=&\frac{e^{-i\ell\phi_2}}{R}
\frac{I_\ell(mr_2)}{I_\ell^2-I_{\ell-1}^2}
\end{eqnarray}
And finally,
\begin{mathletters}\label{eq:G-solution}
\begin{eqnarray}
G_{ss}^{11}(\r_1,\r_2)
&=&
\frac{m}{2\pi}K_0(m|\r_1-\r_2|)
\\
&
\!\!\!+\!\!\!
&\!\!\!\!
\frac{m}{2\pi}\sum_{\ell\in\mathbb{Z}} 
\frac{K_\ell I_\ell+K_{\ell-1} I_{\ell-1}}%
{I_{\ell-1}^2-I_{\ell}^2}\,
I_\ell(mr_1)I_\ell(mr_2)
e^{i\ell(\phi_1-\phi_2)}
\nonumber\\
G_{ss}^{21}(\r_1,\r_2)
&=&
\frac{1}{2\pi R}\sum_{\ell\in\mathbb{Z}} 
\frac{I_\ell(mr_1)I_{\ell+1}(mr_2)}{I_{\ell+1}^2-I_\ell^2}
\ e^{i\ell\phi_1-i(\ell+1)\phi_2}
\\
G_{-ss}^{11}(\r_1,\r_2)&=&
\frac{1}{2\pi R}
\sum_{\ell\in\mathbb{Z}} 
\frac{I_\ell(mr_1)I_\ell(mr_2)}{I_{\ell}^2-I_{\ell-1}^2}
\ e^{i\ell(\phi_1-\phi_2)}
\\
G_{-ss}^{21}(\r_1,\r_2)&=&
\frac{m}{2\pi}\frac{\bar{z}_2-\bar{z}_1}{|\r_1-\r_2|}
K_1(m|\r_1-\r_2|)\\
&&
+\frac{m}{2\pi}
\sum_{\ell\in\mathbb{Z}} 
\frac{K_{\ell} I_{\ell}+K_{\ell-1} I_{\ell-1}}%
{I_{\ell-1}^2-I_\ell^2}
\,
I_{\ell-1}(mr_1) I_{\ell}(mr_2)
e^{i(\ell-1)\phi_1-i\ell\phi_2}
\nonumber
\end{eqnarray}
\end{mathletters}
From equations~(\ref{eq:Gintegral}) it can be seen that the remaining
Green functions can be easily deduced since they obey
\begin{mathletters}\label{eq:symmetry}
\begin{eqnarray}
G_{ss'}^{12}(\r_1,\r_2)&=&\overline{G_{ss'}^{21}(\r_1,\r_2)}\\
G_{ss'}^{22}(\r_1,\r_2)&=&\overline{G_{ss'}^{11}(\r_1,\r_2)}
\end{eqnarray}
Other useful symmetry relations between the Green functions are
\begin{eqnarray}
G_{ss'}^{11}(\r_1,\r_2)=\overline{G_{ss'}^{11}(\r_2,\r_1)}&&
G_{ss'}^{22}(\r_1,\r_2)=\overline{G_{ss'}^{22}(\r_2,\r_1)}\\
G_{ss'}^{21}(\r_1,\r_2)=-G_{ss'}^{21}(\r_2,\r_1)&&
G_{ss'}^{12}(\r_1,\r_2)=-G_{ss'}^{12}(\r_2,\r_1)
\end{eqnarray}
\end{mathletters}

\subsection{Density}

The density is obtained from equation~(\ref{eq:density-def}) and it
reads
\begin{equation}\label{eq:density}
n_{s}(\r)=n_b+\frac{m^2}{2\pi} \sum_{\ell\in\mathbb{Z}}
\frac{K_{\ell}I_{\ell}+I_{\ell-1} 
K_{\ell-1}}{I_{\ell}^2-I_{\ell-1}^2}
\ I_\ell(mr)^2
\end{equation}
where $n_b$ is the bulk density of the infinite system which is
formally divergent when the cutoff vanishes.
Writing formally the bulk density as
\begin{equation}
n_b=\frac{m}{2\pi} \sum_{\ell\in\mathbb{Z}}
I_\ell K_\ell\ ,
\end{equation}
rearranging the terms in equation~(\ref{eq:density}) and using the
Wronskian 
\begin{equation}
I_\ell K_{\ell-1} + I_{\ell-1} K_{\ell}
=
\frac{1}{mR}
\end{equation}
the density at the boundary can be written as
\begin{equation}
n_s(R)=\frac{1}{2\pi R} 
\sum_{\ell\in\mathbb{Z}} \frac{I_\ell I_{\ell-1}}{I_{\ell-1}^2
I_{\ell}^2}
\end{equation}
The above expression clearly vanishes since the term $\ell$ is
canceled by the term $-\ell+1$. So we recover the expected result
\begin{equation}
n_s(R)=0
\end{equation}
The strong repulsion between a charge and its image cause the density
at the boundary to vanish.

%
%
\begin{figure}
\epsfbox{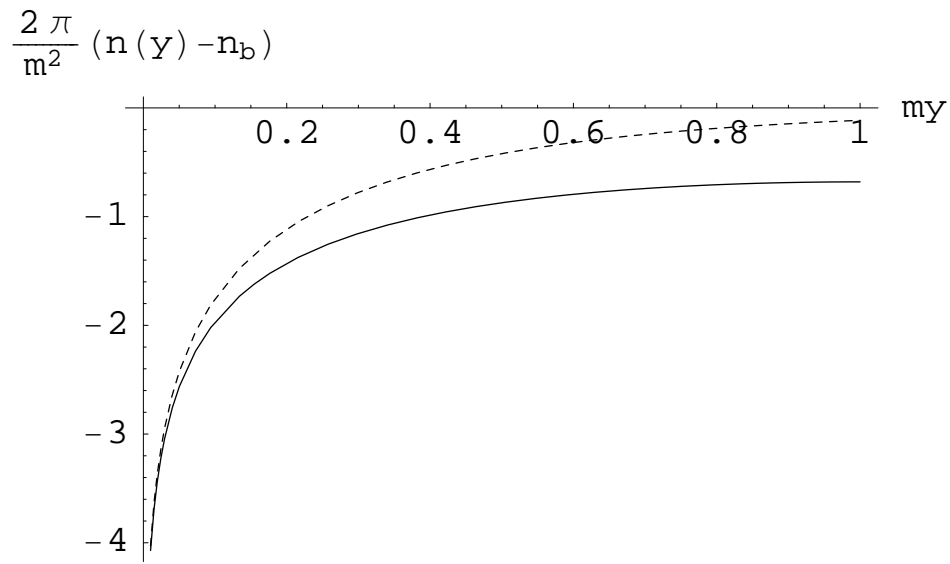}
\caption{\label{fig:density-R1} Difference between the charge density and
the bulk charge density $n_s(y)-n_b$ as a function of the distance
$y=R-r$ from the wall. The dashed curve represents case of a
two-component plasma near an infinite plane wall. The solid curve
represents the density in the disk case with $mR=1$ }
\end{figure}
%
%
%
%
\begin{figure}
\epsfbox{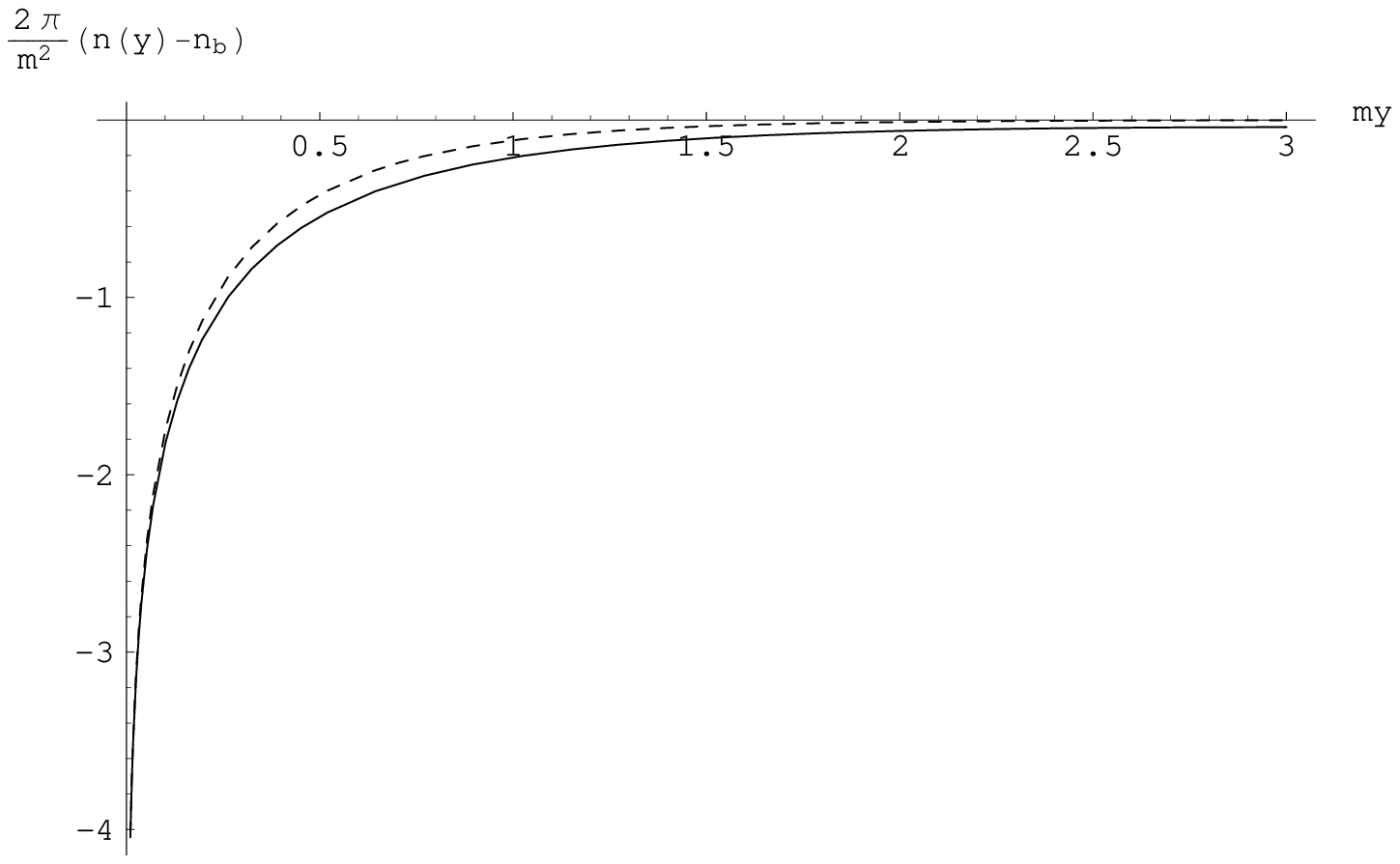}
\caption{\label{fig:density-R3} Difference between the charge density and
the bulk charge density $n_s(y)-n_b$ as a function of the distance
$y=R-r$ from the wall. The dashed curve represents case of a
two-component plasma near an infinite plane wall. The solid curve
represents the density in the disk case with $mR=3$ }
\end{figure}
%
%

In reference~[\cite{JancoSamaj-diel}] it was shown that the density
near an infinite ideal dielectric plane wall is
\begin{equation}\label{eq:density-plain-wall}
n_{s}(y)=n_b-\frac{m^2}{2\pi} K_0(2my)
\end{equation}
where $y$ is the distance from the wall. To compare our result for the
density in a disk and result~(\ref{eq:density-plain-wall}) near an
infinite plane wall we plot in Figures~\ref{fig:density-R1}
and~\ref{fig:density-R3} both densities as a function of the distance
from the wall for disks of different sizes $R=1/m$ and $R=3/m$. In
both cases the density for small distances $y$ behaves as
\begin{equation}
n_s(y)-n_b=\frac{m^2}{2\pi} \ln (2my) + O(1)
\end{equation}
It can also be seen in Figures~\ref{fig:density-R1}
and~\ref{fig:density-R3} that the density decays faster for the
semi-infinite system (plane wall) than in the disk case. This effect
is stronger for the small disk $mR=1$.

\subsection{Correlations}

From the Green functions~(\ref{eq:G-solution}) we obtain the two-body
correlation functions using
equation~(\ref{eq:correlations-def}). Using the symmetry
relations~(\ref{eq:symmetry}) the correlation between a particle of
sign $s$ at $\r_1$ and a particle of sign $s'$ at $\r_2$ reads
\begin{equation}\label{eq:corr-G}
n_{ss'}^{(2)T}(\r_1,\r_2)=
-m^2\left[ \left|G_{ss'}^{11}(\r_1,\r_2)\right|^2
-\left|G_{ss'}^{21}(\r_1,\r_2)\right|^2\right]
\end{equation}
This gives
\begin{mathletters}\label{eq:corr-explicit}
\begin{eqnarray}
n_{ss}^{(2)T}(\r_1,\r_2)&=&
-\frac{m^4}{(2\pi)^2}
\Bigg|K_0(m|\r_1-\r_2|)
\\
&&
\!\!\!+
\sum_{\ell\in\mathbb{Z}} 
\frac{K_\ell I_\ell+K_{\ell-1} I_{\ell-1}}%
{I_{\ell-1}^2-I_{\ell}^2}
I_\ell(mr_1)I_\ell(mr_2)
e^{i\ell(\phi_1-\phi_2)}
\Bigg|^2
\nonumber\\
&&
\!\!\!+
\left(\frac{m}{2\pi R}\right)^2
\left|
\sum_{\ell\in\mathbb{Z}} 
\frac{I_\ell(mr_1)I_{\ell+1}(mr_2)}{I_{\ell+1}^2-I_\ell^2}
\ e^{i\ell\phi_1-i(\ell+1)\phi_2}
\right|^2
\nonumber
\end{eqnarray}
and
\begin{eqnarray}
n_{-ss}^{(2)T}(\r_1,\r_2)&=&
\frac{m^4}{(2\pi)^2}
\Bigg|
\frac{\bar{z}_2-\bar{z}_1}{|\r_1-\r_2|}
K_1(m|\r_1-\r_2|)\\
&&
\!\!\!+
\sum_{\ell\in\mathbb{Z}} 
\frac{K_{\ell} I_{\ell}+K_{\ell-1} I_{\ell-1}}%
{I_{\ell-1}^2-I_\ell^2}
\,
I_{\ell-1}(mr_1) I_{\ell}(mr_2)
e^{i(\ell-1)\phi_1-i\ell\phi_2}
\Bigg|^2
\nonumber\\
&&
\!\!\!
-
\left(\frac{m}{2\pi R}\right)^2
\left|
\sum_{\ell\in\mathbb{Z}} 
\frac{I_\ell(mr_1)I_\ell(mr_2)}{I_{\ell}^2-I_{\ell-1}^2}
\ e^{i\ell(\phi_1-\phi_2)}
\right|^2
\nonumber
\end{eqnarray}
\end{mathletters}
From equation~(\ref{eq:corr-G}) and the boundary
conditions~(\ref{eq:G-bc}) it is clear that if one point is on the
boundary
\begin{equation}
n_{ss'}^{(2)T}(\r_1,\r_2)=0
\qquad
\mathrm{if}
\ \r_1\in\partial D
\ 
\mathrm{or}
\ \r_2\in\partial D
\end{equation}
as expected due to the strong repulsion between a charge and its image.

%
%
\begin{figure}
\epsfbox{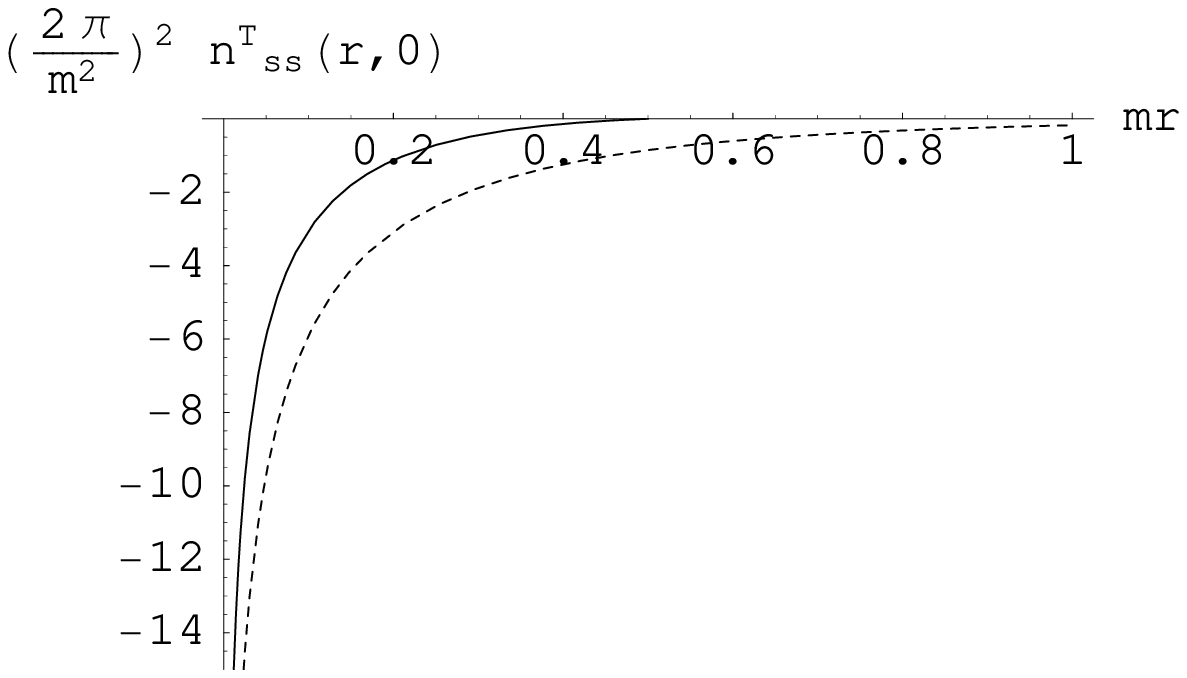}
\caption{\label{fig:corr-pp-R1}
Two-body density $n^{(2)T}_{ss}(\r,0)$ with one point fixed on the
center of the disk for $mR=1$. The dashed curve represents the bulk
correlation and the solid curve the correlation for the disk case.
}
\end{figure}
\begin{figure}
\epsfbox{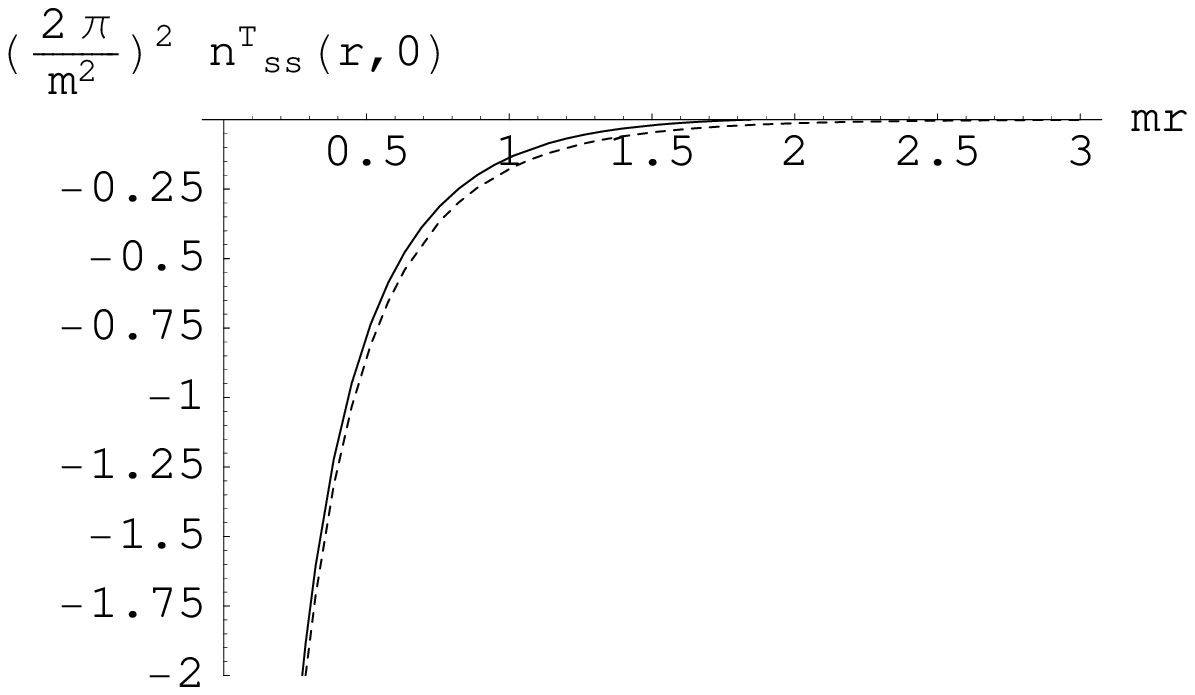}
\caption{\label{fig:corr-pp-R3}
Two-body density $n^{(2)T}_{ss}(\r,0)$ with one point fixed on the
center of the disk for $mR=3$. The dashed curve represents the bulk
correlation and the solid curve the correlation for the disk case.
}
\end{figure}
%
%
%
%
\begin{figure}
\epsfbox{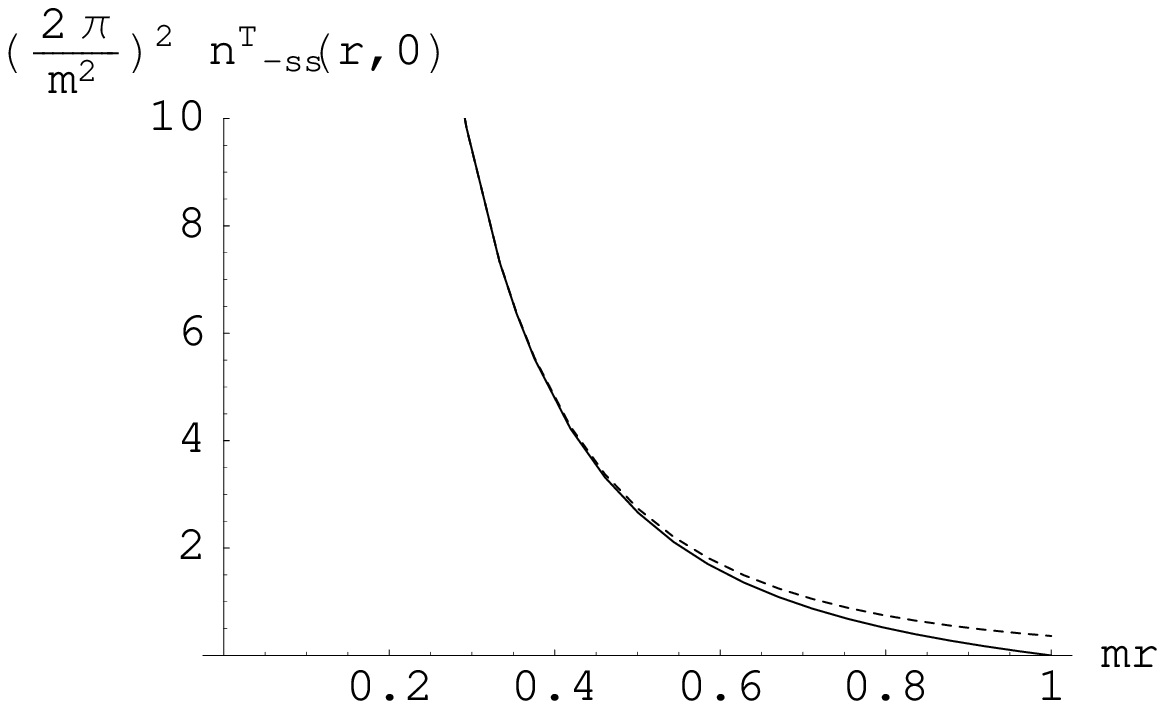}
\caption{\label{fig:corr-pm-R1}
Two-body density $n^{(2)T}_{-ss}(\r,0)$ with one point fixed on the
center of the disk for $mR=1$. The dashed curve represents the bulk
correlation and the solid curve the correlation for the disk case.
}
\end{figure}
\begin{figure}
\epsfbox{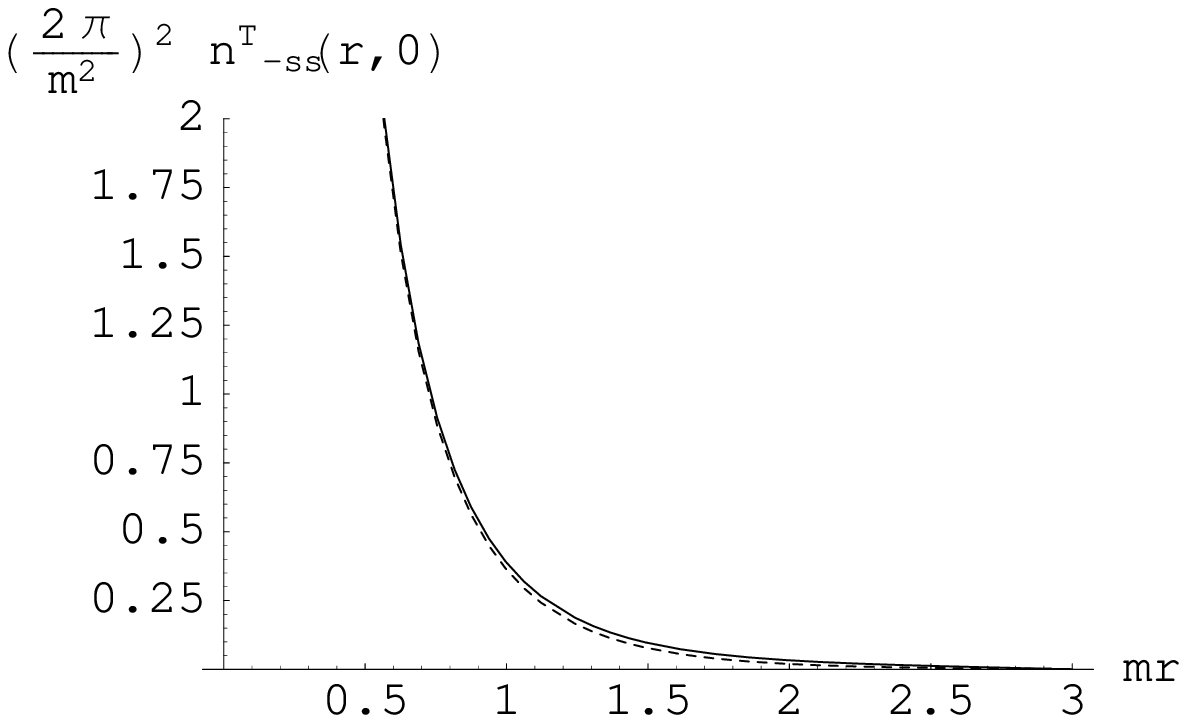}
\caption{\label{fig:corr-pm-R3}
Two-body density $n^{(2)T}_{-ss}(\r,0)$ with one point fixed on the
center of the disk for $mR=3$. The dashed curve represents the bulk
correlation and the solid curve the correlation for the disk case.
}
\end{figure}
%
%

If $\r_2=0$ the above expressions~(\ref{eq:corr-explicit}) simplify
to
\begin{mathletters}\label{eq:corr-r2=0}
\begin{eqnarray}
n_{ss}^{(2)T}(\r,0)&=&
-\left(
\frac{m^2}{2\pi}
\right)^2
\left[
K_0(mr)+
\frac{K_0 I_0+K_1 I_1}{I_1^2-I_0^2}
I_1(mr)
\right]^2
\\
&&
+
\left(\frac{m}{2\pi R}\right)^2
\frac{I_1(mr)^2}{(I_1^2-I_0^2)^2}
\nonumber\\
n_{-ss}^{(2)T}(\r,0)&=&
\left(
\frac{m^2}{2\pi}
\right)^2
\left[
-K_1(mr)+
\frac{K_0 I_0+K_1 I_1}{I_1^2-I_0^2}
I_1(mr)
\right]^2\\
&&
-
\left(\frac{m}{2\pi R}\right)^2
\frac{I_0(mr)^2}{(I_1^2-I_0^2)^2}
\nonumber
\end{eqnarray}
\end{mathletters}
It is interesting to compare these expressions with the bulk
correlations for an infinite system\citenote{CornuJanco}
\begin{mathletters}
\begin{eqnarray}
n_{ss,\,\mathrm{bulk}}^{(2)T}(\r)&=&-\left(\frac{m^2}{2\pi}\right)^2
\left[K_0(mr)\right]^2\\
n_{-ss\,\mathrm{bulk}}^{(2)T}(\r)&=&\left(\frac{m^2}{2\pi}\right)^2
\left[K_1(mr)\right]^2
\end{eqnarray}
\end{mathletters}
Figures~\ref{fig:corr-pp-R1} and~\ref{fig:corr-pp-R3} show the
two-body density $n_{ss}^{(2)T}(\r,0)$ for particles of same sign
compared to the bulk values for different values of $R$ and
Figures~\ref{fig:corr-pm-R1} and~\ref{fig:corr-pm-R3} show the
two-body density $n_{-ss}^{(2)T}(\r,0)$ for particles of different
sign. For a small disk with $mR=1$ there is a notable difference. In
the disk case the correlations decay faster than in the bulk. This can
be easily understood since there is a strong repulsion between a
particle and the boundary. But this difference can be hardly noted if
the disk is larger. For $mR=3$ it can be seen in
Figure~\ref{fig:corr-pp-R3} that the difference between the bulk and
the disk case is very small (notice the change of scale in the
vertical axis between Figures~\ref{fig:corr-pp-R1}
and~\ref{fig:corr-pp-R3}).

%
%
\begin{figure}
\epsfbox{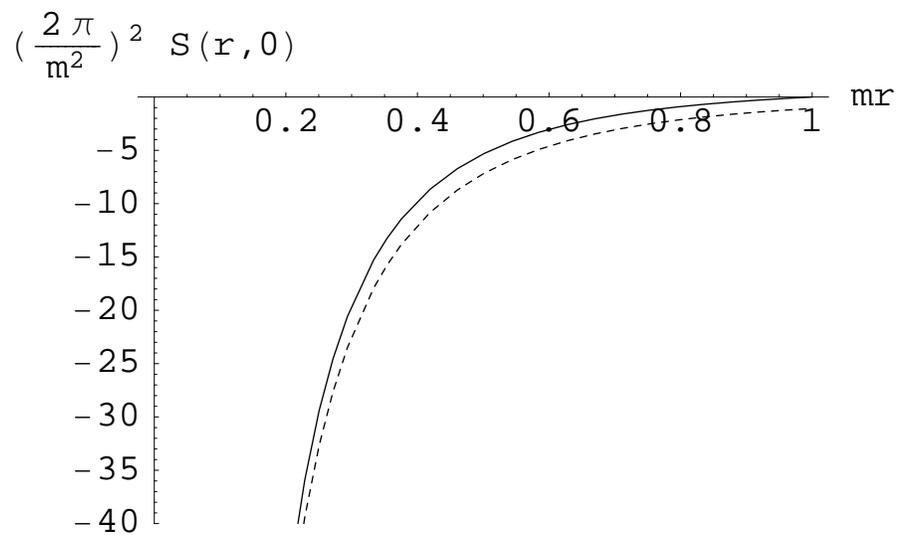}
\caption{\label{fig:struc-R1}
Structure function $S(\r,0)$ with one point fixed on the
center of the disk for $mR=1$. The dashed curve represents the bulk
correlation and the solid curve the correlation for the disk case.
}
\end{figure}
\begin{figure}
\epsfbox{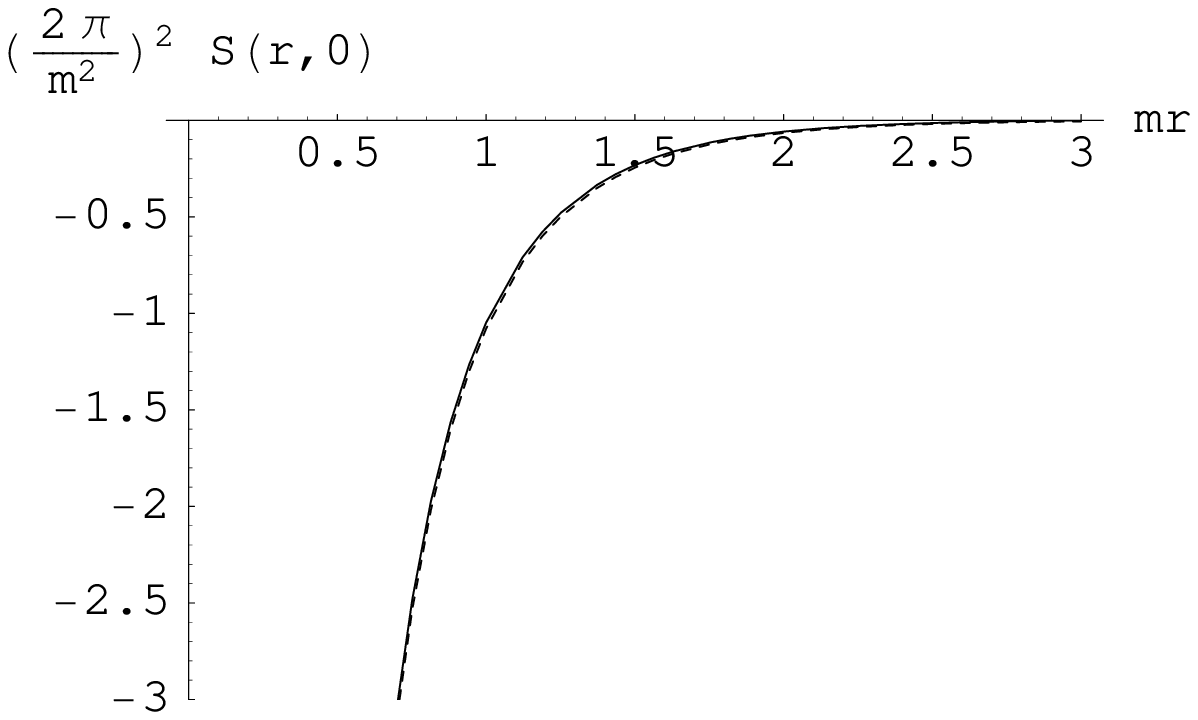}
\caption{\label{fig:struc-R3}
Structure function $S(\r,0)$ with one point fixed on the
center of the disk for $mR=3$. The dashed curve represents the bulk
correlation and the solid curve the correlation for the disk case.
}
\end{figure}
%
%

In Figures~\ref{fig:struc-R1} and~\ref{fig:struc-R3}
we plot the structure function (charge-charge correlation)
\begin{equation}
S(\r_1,\r_2)=2(n_{ss}^{(2)T}(\r_1,\r_2)
-n_{-ss}^{(2)T}(\r_1,\r_2))
\end{equation}
with one point fixed at center of the disk $\r_2=0$. For the small
disk $mR=1$ there is a clear difference between the bulk case and the
disk case. Due to the repulsion between a particle and its image, the
screening cloud is more concentrated in the center of the disk than in
the bulk case. But for the large disk $mR=3$ the difference is hardly
noticeable. Notice again the change of scale between
Figures~\ref{fig:struc-R1} and~\ref{fig:struc-R3}. It is interesting
to note that there is not much difference between the bulk and the
disk case if the radius of the disk is a few orders the screening
length $m^{-1}$ and larger. We only notice differences when $R\sim
m^{-1}$ and smaller.

\section{Conclusion}

We have studied the two-component plasma with coupling constant
$\Gamma=2$ confined in a disk of radius $R$ with ideal dielectric
boundaries: the electric potential obeys Neumann boundary
conditions. The model is solvable by using a mapping of the Coulomb
system into a four-component free Fermi field. We have computed the
grand-potential, the density and correlation functions.

The grand-potential can be formally written as an average of the
grand-potential for ideal conductor boundaries and the same
grand-potential for ideal conductor boundaries but with the sign of
the fugacity changed. For ideal conductor boundaries the surface
tension is infinite when the cutoff vanishes. Here, the average makes
the surface tension finite. This fact also appears for a two-component
plasma in a strip.\citenote{JancoSamaj-diel}

The Neumann boundary conditions for the electric potential being
conformally invariant it is expected that the grand-potential of the
system exhibits a universal finite-size correction $(1/6)\ln R$. This
was explicitly checked on this solvable model. We also checked this
universal finite-size correction on the model of the one-component
plasma at $\Gamma=2$ in the same confined geometry which was solved
some time ago.\citenote{Smith}

The density vanishes for a point on the boundary of the disk. This is
expected since there is a strong repulsion between the particles and
the boundary due to the image forces. This is also true for the
correlations, they vanish if one point is on the boundary. We compared
the correlations functions for an infinite system without boundaries
and the present system in a disk. Due to the repulsion between the
particles and the boundary, the screening cloud around a particle in
the center of the disk is smaller than the one for an infinite
system. But the difference between the screening clouds is only
noticeable for disks with radius of the order of the screening length
and smaller. If the disk has a radius a few orders larger than the
screening length, the difference can be hardly noted.

\section*{Acknowledgements}

The author would like to thank B.~Jancovici for stimulating
discussions, useful comments and for a critical reading of the
manuscript.  Partial financial support from COLCIENCIAS and BID
through project \# 1204-05-10078 is acknowledged.

\end{document}